\documentclass[12pt,final,oneside,onecolumn]{IEEEtran}
\usepackage{color}
\usepackage[utf8]{inputenc}
\usepackage[mathscr]{euscript}
\usepackage{amssymb}
\usepackage{amsfonts}
\usepackage{filecontents}
\usepackage{amsfonts}
\usepackage{filecontents}
\usepackage{bm}
\usepackage{amsthm}
\usepackage{algorithm}
\usepackage{multicol}
\usepackage{setspace}
\usepackage[noend]{algpseudocode}
\newtheoremstyle{exampstyle}
{3pt} % Space above
{2pt} % Space below
{} % Body font
{} % Indent amount
{\bfseries} % Theorem head font
{.} % Punctuation after theorem head
{.5em} % Space after theorem head
{} % Theorem head spec (can be left empty, meaning `normal')
\theoremstyle{exampstyle}

\newtheorem{theo}{Theorem}
\newtheorem{prep}{Proposition}
\newtheorem{corol}{Corollary}
\theoremstyle{definition}

\newtheorem{assum}{Assumption}
\theoremstyle{remark}
\newtheorem{remark}{Remark}

\ifCLASSOPTIONcompsoc
  \usepackage[nocompress]{cite}
\else
  \usepackage{cite}
\fi

\ifCLASSINFOpdf
   \usepackage[pdftex]{graphicx}
\else
  \usepackage[dvips]{graphicx}
\fi

\usepackage{amsmath}

\usepackage{array}
\usepackage{mdwmath}
\usepackage{mdwtab}
\usepackage{eqparbox}
\ifCLASSOPTIONcompsoc
  \usepackage[caption=false,font=footnotesize,labelfont=sf,textfont=sf]{subfig}
\else
  \usepackage[caption=false,font=footnotesize]{subfig}
\fi

\begin{document}

\title{Analysis of Distributed ADMM Algorithm for Consensus Optimization in Presence of Node Error}
\author{
\IEEEauthorblockN{Layla Majzoobi$^+$, Farshad Lahouti$^{*}$, and Vahid~Shah-Mansouri$^+$}

\IEEEauthorblockA{+School of Electrical and Computer Engineering, University of Tehran\\
*Electrical Engineering Department, California Institute of Technology}}

\IEEEtitleabstractindextext{%

\begin{abstract}
Alternating Direction Method of Multipliers (ADMM) is a popular convex optimization algorithm, which can be employed for solving distributed consensus optimization problems. In this setting agents locally estimate the optimal solution of an optimization problem and exchange messages with their neighbors over a connected network. The distributed algorithms are typically exposed to different types of errors in practice, e.g., due to quantization or communication noise or loss. We here focus on analyzing the convergence of distributed ADMM for consensus optimization in presence of additive random node error, in which case, the nodes communicate a noisy version of their latest estimate of the solution to their neighbors in each iteration. We present analytical upper and lower bounds on the mean squared steady state error of the algorithm in case that the local objective functions are strongly convex and have Lipschitz continuous gradients. In addition we show that, when the local objective functions are convex and the additive node error is bounded, the estimation error of the noisy ADMM for consensus optimization is also bounded. Numerical results are provided which demonstrate the effectiveness of the presented analyses and shed light on the role of the system and network parameters on performance.
\end{abstract}

\begin{IEEEkeywords}
ADMM algorithm, Consensus optimization, Convergence, Node error, Steady State Error.
\end{IEEEkeywords}}

\maketitle

\IEEEdisplaynontitleabstractindextext

\IEEEpeerreviewmaketitle

\ifCLASSOPTIONcompsoc
\IEEEraisesectionheading{\section{Introduction}\label{sec:introduction}}
\else
\section{Introduction}
\label{sec:introduction}
\let\thefootnote\relax\footnote{Preliminary results on this research has been reported in IEEE International Conference on Acoustics, Speech and Signal Processing (ICASSP), China, 2016 \cite{majzoobi2016analysis}.} In the advent of big data, the Internet of Things and cyber-physical systems, distributed processing algorithms have attracted increasing research attention. Considerable problems in these areas including network flow control, feature extraction and power system state estimation, are formulated as convex optimization problems, and use distributed optimization methods
\cite{low1999optimization}-\nocite{erseghe2014distributed,rabbat2004distributed,nedic2009distributed,johansson2009randomized,ghadimi2013multi,li2013distributed}\cite{chang2014distributed}.
These algorithms may be characterized by their range of applicability, convergence behavior, performance, and computational complexity.

Alternating Direction Method of Multipliers (ADMM) is one of the popular convex optimization algorithms which can be implemented in a distributed manner. Wide range of applicability and convergence with an adequate accuracy within a few tens of iterations make ADMM very useful in practice. {\color{black}Different applications and convergence behavior of ADMM algorithm are discussed in~\cite{boyd2011distributed}\nocite{shen2012distributed}\nocite{lin2013design}\nocite{liu2013tensor}
\nocite{danaher2014joint}\nocite{qing2014robust}\nocite{deng2012global}-\nocite{he20121}\nocite{deng2014parallel}
\nocite{goldstein2014fast}\cite{giselsson2017linear}. An upper bound on the global linear convergence rate of ADMM algorithm for strongly convex objective function under smoothness assumptions is provided in \cite{giselsson2017linear}. A parameter selection method which optimizes the derived convergence bound is also presented in this work.}

Decentralized consensus optimization is an important family of optimization problems which is formulated as
\begin{equation}\label{consProblem}
\underset{\tilde{\mathbf{x}}}{\operatorname{min}}\quad\sum_{i=1}^{N} f_i(\tilde{\mathbf{x}})
\end{equation}
where $\tilde{\textbf{x}} \in \mathbb{R}^n$ and $f_i(\tilde{\mathbf{x}}) : \mathbb{R}^n \rightarrow \mathbb{R}$ are convex functions. $N$ agents aim to optimize the sum of local objective functions, $f_i(\tilde{\mathbf{x}}), i\in\{1,...,N\}$, over a global variable $\tilde{\mathbf{x}}$. {\color{black}Distributed ADMM-based algorithms for consensus optimization problems can be classified into two categories. In the first one, there is a central collector or fusion center that all agents communicate with. Indeed the agents and the fusion center form a star network \cite{boyd2011distributed}. In the second one, there is not a fusion center, and the underlying network between the agents can have any connected topology ~\cite{mateos2010distributed}\nocite{etemad2016resilient}\nocite{mota2013d}-\cite{Wei2014}. While the former category has better convergence rate, the flexibility of network topology in the latter one is more appealing.
A framework is provided in \cite{ma2018fast}, which improves the convergence property of the latter category by allowing multiple fusion centers in the network.}

 The distributed optimization algorithms rely on local computations and communication between neighbors. They potentially suffer from different types of errors and imperfections such as computation, quantization and communication errors and link failure which could substantially affect the performance of  distributed  optimization methods. The subgradient-push algorithm which is convergent over time varying networks is proposed in \cite{nedic2015distributed}. ADMM-based algorithms which are resilient to link failure are proposed in \cite{majzoobi2018analysis}. {\color{black}Certain local computations are too complex to be carried out exactly and are usually replaced by approximations. Inexact variants of ADMM algorithm are  proposed  in \cite{mateos2010distributed,chang2015multi,ling2015dlm,mokhtari2015decentralized}. In \cite{mateos2010distributed}, three distributed ADMM-based algorithms are proposed to solve the Lasso problem offering different computational complexity and convergence rate trade-offs. In \cite{chang2015multi} each ADMM update step, which includes solving an optimization subproblem, is replaced by an inexact  step which reduces the computational cost of the algorithm by an order of magnitude. In \cite{ling2015dlm} and \cite{mokhtari2015decentralized}, the Lagrangian minimization in each ADMM step is, respectively, replaced by linear and  quadratic approximations of the objective function. The quadratic approximation is more complex than the linear approximation, but results in an algorithm with better convergence properties.} {\color{black}A distributed subgradient method for consensus optimization which is convergent in presence of quantization error is proposed in \cite{nedic2008distributed}. In \cite{schizas2008consensus}, ADMM-based distributed algorithms are proposed for parameter estimation in ad hoc wireless sensor networks which exhibit resilience in the presence of communication/quantization noise. For the least square objective function, the steady state error of the algorithm is analyzed. A robust ADMM-based distributed algorithm for average consensus problem in presence of noise is presented in \cite{erseghe2011fast}.

In this work, we focus on analyzing the convergence behavior of distributed ADMM for consensus optimization over a connected network in presence of additive node error. We consider general convex or strongly convex local objective functions.} In this setting, the latest estimate of the solution computed at each node is subject to additive random noise prior to transmission to neighboring nodes (this for instance could be due to quantization error \cite{marco2005validity}). We provide analytical upper and lower bounds on the mean squared steady state error of the noisy ADMM algorithm for consensus, when local objective functions are strongly convex and have Lipschitz continuous gradients. In addition, we show that for convex, proper and closed local objective functions and in presence of bounded node error, the steady state error of noisy ADMM algorithm for consensus is bounded. Extensive numerical results are provided which shows the effectiveness of the presented analyses. We also study the effect of different factors, such as noise variance, network topology and ADMM algorithm parameter on the provided bounds. We also propose a method to tune the algorithm parameter, which results in a faster convergence rate and a smaller steady state error.

The rest of the paper is organized as follows. The system model and prior art is reviewed in good details in Section \ref{sec:systemModel}. Section \ref{sec:convAnal} analyzes the convergence behavior of the algorithm in presence of additive node error. Analytical results are numerically assessed in Section \ref{sec:numerical}, and conclusions are drawn in Section \ref{sec:concl}.

%*************Algorithm 1
\begin{algorithm*}[t]
	\caption{: Distributed ADMM-based algorithm for consensus optimization problem}\label{Alg1}
	Input functions $f_i$; Initialization: for all $i \in \mathcal{V}$, set $\mathbf{x}_i^0 = \boldsymbol{\alpha}_i^0= 0_{n\times1} \in \mathbb{R}^n$; Set algorithm parameter $ c > 0$; $ k = 0$:
	\begin{algorithmic}
		\ForAll{$k = 0, 1, \dots, $ every node $i$}
		\State 1. Update $\mathbf{x}_i^{k+1}$ by solving $\nabla f_i (\mathbf{x}_i ^{k+1}) + \boldsymbol{\alpha}_i^k + 2c|\mathcal{N}_i|\mathbf{x}_i^{k+1} - c(|\mathcal{N}_i|\mathbf{x}_i^k + \sum_{j \in \mathcal{N}_i} \mathbf{x}_j^k) = \mathbf{0}$ ;
		\State 2. Node $i$ sends $\mathbf{x}_i^{k+1}$ to its neighbors;
		\State 3. Update $\boldsymbol{\alpha}_i^{k+1} = \boldsymbol{\alpha}_i^k + c(|\mathcal{N}_i| \mathbf{x}_i^{k+1}- \sum_{j \in \mathcal{N}_i} \mathbf{x}_j^{k+1} )$;
		\EndFor
	\end{algorithmic}
\end{algorithm*}
%***********
%***********************SECTION1
\section{Preliminaries}
\label{sec:systemModel}
Consider a set of $N$ nodes that intend to solve the consensus optimization problem in (\ref{consProblem}) in a distributed manner and over a given connected network with $E$ undirected links. The network can be represented by the graph $\mathcal{G} = (\mathcal{V}, \mathcal{A})$, where $\mathcal{V}$ is the set of nodes, $|\mathcal{V}| = N$,  and $\mathcal{A}$ is the set of arcs, $|\mathcal{A}| = 2E$. Communications of nodes are synchronous and restricted to the neighbors.

In this section, we briefly review the ADMM algorithm and the known convergence results from \cite{Wei2014} for completeness. To apply the ADMM algorithm to problem in (\ref{consProblem}), we reformulate the problem as
\begin{equation}\label{admmForm}
\begin{split}
\underset{\mathbf{x}_i , \mathbf{z}_{ij}}{\operatorname{min}}\quad&\sum_{i=1}^{N} f_i(\mathbf{x}_i) \\
\text{s.t.} \quad & \mathbf{x}_i = \mathbf{z}_{ij},\quad \mathbf{x}_j= \mathbf{z}_{ij} \quad \forall(i,j) \in \mathcal{A}
\end{split}
\end{equation}
\noindent where $\mathbf{x}_i \in \mathbb{R}^n$  is the copy of the global optimization variable $\tilde{\mathbf{x}}\in \mathbb{R}^n$ at node $i$ and $\mathbf{z}_{ij}\in \mathbb{R}^n$ is the variable, which enforces equality of $\mathbf{x}_i$ and $\mathbf{x}_j$ at the optimal point. Concatenating $\mathbf{x}_i$s and $\mathbf{z}_{ij}$s respectively into $\mathbf{x} \in \mathbb{R}^{nN}$ and $\mathbf{z} \in \mathbb{R}^{2nE}$, (\ref{admmForm}) can be rewritten in the following form
\begin{equation}
\label{matrixForm}
\underset{\mathbf{x}, \mathbf{z}}{\operatorname{min}}\quad f(\mathbf{x}) + g(\mathbf{z});\hspace{0.5cm}
\text{s.t.}\quad\mathbf{Ax} + \mathbf{Bz} = 0
\end{equation}
\noindent where $f(\mathbf{x}) = \sum_{i=1}^{N} f_i(\mathbf{x}_i)$, $g(\mathbf{z}) = 0$. Matrix $\mathbf{A}$ can be partitioned as  $\mathbf{A} = [\mathbf{A}_1;\mathbf{A}_2]$, where $\mathbf{A}_1 = \hat{\mathbf{A}}_1 \otimes  \boldsymbol{I}_n$ and $\mathbf{A}_2 = \hat{\mathbf{A}}_2 \otimes  \boldsymbol{I}_n$, and $\otimes$ represents the Kronecker product. Elements of $\hat{\mathbf{A}}_1$ and $\hat{\mathbf{A}}_2$  are zero or one.
Assuming that $(i,j) \in \mathcal{A}$ is an arc in the network graph, the $(m,i)$ element of $\hat{\mathbf{A}}_1$ and the $(m,j)$ element of $\hat{\mathbf{A}}_2$ are one if  $\mathbf{z}_{ij}$ is the $m^{th}$ block of $\mathbf{z}$, $\forall \hspace{1mm} m \in [1, 2E]$. Also, we have $\mathbf{B} = [-\boldsymbol{I}_{2nE};-\boldsymbol{I}_{2nE}]$. Forming augmented Lagrangian and applying ADMM algorithm, we have
\begin{subequations}\label{admmIteration}
	\begin{equation}\label{admmA}
	\nabla f(\mathbf{x}^{k+1}) + \mathbf{A}^T\boldsymbol{\lambda}^k + c\mathbf{A}^T(\mathbf{A}\mathbf{x}^{k+1} + \mathbf{B}\mathbf{z}^k) = \mathbf{0}
	\end{equation}
	\begin{equation}\label{admmB}
	\mathbf{B}^T\boldsymbol{\lambda}^k + c\mathbf{B}^T(\mathbf{A}\mathbf{x}^{k+1} + \mathbf{B}\mathbf{z}^{k+1}) = \mathbf{0}
	\end{equation}
	\begin{equation}\label{admmC}
	\boldsymbol{\lambda}^{k+1} - \boldsymbol{\lambda}^k - c(\mathbf{A}\mathbf{x}^{k+1} + \mathbf{B}\mathbf{z}^{k+1}) = \mathbf{0}
	\end{equation}
\end{subequations}
\noindent where $\boldsymbol{\lambda}$ and $c$ are Lagrange multiplier and ADMM parameter, respectively; $\nabla f(\mathbf{x}^{k+1})$ is gradient or
subgradient of $f(\mathbf{x})$ at $\mathbf{x}^{k+1}$ depending on the differentiability of $f$. Considering $\boldsymbol{\lambda} = [\boldsymbol{\beta};\boldsymbol{\nu}]$ with $\boldsymbol{\beta},\boldsymbol{\nu} \in \mathbb{R}^{2nE}$, $\mathbf{M}_+ = \mathbf{A}_1^T + \mathbf{A}_2^T$ and $\mathbf{M}_- = \mathbf{A}_1^T - \mathbf{A}_2^T$, it can be shown that $\boldsymbol{\beta} = - \boldsymbol{\nu}$ and with some
algebraic manipulations \cite{Wei2014}, we have
\begin{subequations}\label{midForm}
	\begin{equation}\label{midFormA}
	\nabla f(\mathbf{x}^{k+1}) + \mathbf{M}_-\boldsymbol{\beta}^{k+1} - c\mathbf{M}_+(\mathbf{z}^k - \mathbf{z}^{k+1}) = \mathbf{0}
	\end{equation}
	\begin{equation}\label{midFormB}
	\boldsymbol{\beta}^{k+1} - \boldsymbol{\beta}^k - \frac{c}{2}\mathbf{M}_-^T\mathbf{x}^{k+1} = \mathbf{0}
	\end{equation}
	\begin{equation}\label{midFormC}
	\frac{1}{2}\mathbf{M}_+^T\mathbf{x}^{k+1} - \mathbf{z}^{k+1} = \mathbf{0}
	\end{equation}
\end{subequations}
Multiplying \eqref{midFormB} by $\mathbf{M}_-$ and substituting the result in \eqref{midFormA}, and also replacing \eqref{midFormC} in \eqref{midFormA}, we have
\begin{subequations}\label{eq:twoStep}
	\begin{equation}\label{eq:first}
	\nabla f(\mathbf{x}^{k+1}) + \mathbf{M}_-\boldsymbol{\beta}^{k} + 2c\mathbf{W}\mathbf{x}^{k+1} - \frac{c}{2}\mathbf{M}_+\mathbf{M}_+^T\mathbf{x}^k = \mathbf{0}
	\end{equation}
	\begin{equation}\label{eq:sec}
	\boldsymbol{\beta}^{k+1} - \boldsymbol{\beta}^k - \frac{c}{2}\mathbf{M}_-^T\mathbf{x}^{k+1} = \mathbf{0},
	\end{equation}	
\end{subequations}
 where
		\begin{equation}\label{eq:wDef}
		\mathbf{W} = \frac{1}{4}\big(\mathbf{M}_+\mathbf{M}_+^T + \mathbf{M}_-\mathbf{M}_-^T\big),
		\end{equation}
is the extended degree matrix of the underlying network. By “extended”, we mean replacing every $1$ by $\mathbf{I}_n$, and $0$  by $\mathbf{0}_n$ in the original definition of this matrix \cite{Wei2014}. Multiplying two sides of \eqref{eq:sec} by $\mathbf{M}_-$, and
defining
$\boldsymbol{\alpha} = [\boldsymbol{\alpha}_1;  \boldsymbol{\alpha}_2;\dots;\boldsymbol{\alpha}_N]=\mathbf{M}_-\boldsymbol{\beta} \in \mathbb{R}^{nN}$,
a relatively simple fully decentralized algorithm is obtained in \cite{ Wei2014} and presented in Algorithm \ref{Alg1}. In this Algorithm, $\mathcal{N}_i$ is the set of neighbors of node $i$.

It is shown that with the following assumption, the proposed algorithm converges R-linearly to its optimal point over a fixed connected underlying network\cite{Wei2014}.

\begin{assum}\label{assum1}
	 The local objective functions are strongly convex and have Lipschitz continuous gradients. For each agent $i$, and given any $\mathbf{x}_a , \mathbf{x}_b \in \mathbf{R}^n$, we have
\begin{equation*}
\begin{split}
   &\big<\nabla f_i(\mathbf{x}_a) - \nabla f_i(\mathbf{x}_b), \mathbf{x}_a - \mathbf{x}_b\big> \geq m_{f_i}\|\mathbf{x}_a - \mathbf{x}_b\|_2^2,\\
   & \|\nabla f_i(\mathbf{x}_a) - \nabla f_i(\mathbf{x}_b)\|_2 \leq M_{f_i}\|\mathbf{x}_a - \mathbf{x}_b\|_2,
   \end{split}
\end{equation*}
where $m_{f_i} > 0$ and $M_{f_i} > 0$. 
\end{assum}
According to Assumption \ref{assum1}, $f(\mathbf{x})$ is a strongly convex function with module $m_{f} = \min _i m_{f_i}$ and its gradient is Lipschitz continuous with constant $M_f = \max _i M_{f_i}$.

\begin{theo}\label{theoWeiShi}
	Consider the ADMM iterations in (\ref{midForm}), that solves the optimization problem in (\ref{matrixForm}). The primal variables $\mathbf{x}$ and $\mathbf{z}$,
	respectively have their optimal values $\mathbf{x}^*$ and $\mathbf{z}^*$, and the optimal value for the dual variable $\boldsymbol{\beta}$ is
	$\boldsymbol{\beta}^*$. Considering
	\begin{equation}\label{uGDef}
	\mathbf{u}^k \triangleq
	\begin{pmatrix}
	\mathbf{z}^k\\
	\boldsymbol{\beta}^k
	\end{pmatrix},\quad
	\mathbf{u}^* \triangleq
	\begin{pmatrix}
	\mathbf{z}^*\\
	\boldsymbol{\beta}^*
	\end{pmatrix},\quad
	\mathbf{G} \triangleq
	\begin{pmatrix}
	c\mathbf{I}_{2nE} &  0_{2nE}\\
	0_{2nE} & \frac{1}{c}\mathbf{I}_{2nE}
	\end{pmatrix},
	\end{equation}
	if the local objective functions satisfy Assumption \ref{assum1}, and the dual variable $\boldsymbol{\beta}$ is initialized such that
	$\boldsymbol{\beta}^0$ lies in the column space of $\mathbf{M}_-^T$, then for any $\mu > 1$, $\mathbf{u}^{k}$ is Q-linearly convergent to its optimal
	point $\mathbf{u}^*$ with respect to the $G$-norm
	\begin{equation}\label{convergence}
	\|\mathbf{u}^{k+1} - \mathbf{u}^*\|_G^2 \leq \rho \|\mathbf{u}^k - \mathbf{u}^*\|_G^2
	\end{equation}\normalsize
	\noindent where $\rho$ is the convergence rate of the algorithm and \small
	\begin{equation}\label{eq:rho}
	\rho = \frac{1}{1+\delta},
	\end{equation}
	where
	\begin{equation}\label{delta}
	\delta = \min \bigg\{\frac{(\mu - 1)\tilde{\sigma}_{min}^2(\mathbf{M}_-)}{\mu\sigma_{max}^2(\mathbf{M}_+)},\frac{m_f}{\frac{c}{4}\sigma_{max}^2(\mathbf{M}_+) + \frac{\mu}{c}M_f^2\tilde{\sigma}_{min}^{-2}(\mathbf{M}_-)}\bigg\}
	\end{equation}\normalsize
	where $\sigma_{max}(\mathbf{M}_+)$ and $\tilde{\sigma}_{min}(\mathbf{M}_-)$ are the largest and smallest non-zero singular values of $\mathbf{M}_+$
	and $\mathbf{M}_-$, respectively. Further $\mathbf{x}^k$ is R-linearly convergent to $\mathbf{x}^*$, which follows from\small
	\begin{equation}\label{xConverg}
	\|\mathbf{x}^{k+1} - \mathbf{x}^*\|_2^2 \leq \frac{1}{m_f}\|\mathbf{u}^k - \mathbf{u}^*\|_G^2.
	\end{equation}
\end{theo}
\begin{IEEEproof}
	See \cite{Wei2014}.
\end{IEEEproof}
{\color{black}As shown in \cite{Wei2014}, $\mu$ and $c$ parameters which maximize $\delta$ are
\begin{equation}\label{eq:muOpt}
\mu^* = \bigg(1 + \frac{K_G^2}{2K_f^2} - \frac{K_G}{2K_f}\sqrt{\frac{K_G^2}{K_f^2} + 4}\bigg)^{-1},
\end{equation}
\begin{equation}\label{eq:cOpt}
c^* = \frac{2\sqrt{\mu^*}M_f}{\sigma_{max}(\mathbf{M}_+)\tilde{\sigma}_{min}(\mathbf{M}_-)},
\end{equation}
where
\begin{equation*}\label{eq:KG}
K_G = \frac{\sigma_{max}(\mathbf{M}_+)}{\tilde{\sigma}_{min}(\mathbf{M}_-)} , K_f = \frac{M_f}{m_f},
\end{equation*}
and the corresponding $\delta$ is
\begin{equation}\label{eq:de;taOpt}
\delta^* = \frac{1}{2K_f}\sqrt{\frac{1}{K_f^2} + \frac{4}{K_G^2}} - \frac{1}{2K_f^2}.
\end{equation}
Note that since $\rho = \frac{1}{1 + \delta}$ is the upper bound of the convergence rate, $c^*$ does not neccesserily maximize the algorithm convergence rate, but as it is shown in \cite[Figure 3]{Wei2014}, this can accelerate the algorithm. }

As described and according to Algorithm \ref{Alg1}, the nodes performing local processing in each iteration send their last estimation of the optimal point to their neighbors. In practical situations, local processing suffer from different sources of errors, such as truncation and quantization, which can  potentially change the convergence behavior of the algorithm. In the next section, we analyze the performance of Algorithm \ref{Alg1} in presence of an additive random error at the nodes.

%*****************Section2
\section{Distributed ADMM algorithm in presence of node error}
\label{sec:convAnal}
%\subsection{Theoretical Results}
In this section, we consider the scenario where the neighboring nodes only exchange a noisy version of their messages in Step 2 of Algorithm \ref{Alg1}. We model the node error $\mathbf{e}_{x,i}^k$ at node $i$ in iteration $k$, as zero mean, additive and independent, identically distributed (i.i.d) error. We have
	\begin{equation}\label{noisy}
	{\hat {\mathbf x}}^k_i = \mathbf{x}^k_i + \mathbf{e}_{x,i}^k,\hspace{1cm}
	\hat{\mathbf{x}}^k = \mathbf{x}^k + \mathbf{e}_x^k,
	\end{equation}
	where $\hat{\mathbf{x}}_i^k$ is the message that node $i \in \mathcal{N}$ communicates to its neighbors at iteration $k$; and $\hat{\mathbf{x}} \in \mathbb{R}^{nN}$
	and $\mathbf{e}_x^k \in \mathbb{R}^{nN}$ denote the concatenated versions of $\hat{\mathbf{x}}_{i}^k$ and $\mathbf{e}_{x,i}^k$ for all $i$. %The Algorithm 2 presents the proposed distributed optimization method in presence of node error.
%%%%%%%%%%%%%%%%%%%
%********* Algorithm2
\begin{algorithm*}[t]
	\caption{: Distributed ADMM-based algorithm for consensus optimization problem in presence of node error}\label{Alg2}
	Input functions $f_i$; Initialization: for all $i \in \mathcal{V}$, set $\hat{\mathbf{x}}_i^0 = \hat{\boldsymbol{\alpha}}_i^0= 0_{n\times1} \in \mathbb{R}^n$; Set algorithm parameter $ c > 0$; $ k = 0$:
	\begin{algorithmic}
		\ForAll{$k = 0, 1, \dots, $ every node $i$}
		\State 1. Update $\mathbf{x}_i^{k+1}$ by solving $\nabla f_i (\mathbf{x}_i ^{k+1}) + \hat{\boldsymbol{\alpha}}_i^k + 2c|\mathcal{N}_i|\mathbf{x}_i^{k+1} - c(|\mathcal{N}_i|\hat{\mathbf{x}}_i^k + \sum_{j \in \mathcal{N}_i} \hat{\mathbf{x}}_j^k) = \mathbf{0}$ ;
		\State 2. Node $i$ sends $\hat{\mathbf{x}}_i^{k+1} = \mathbf{x}_i^{k+1} + \mathbf{e}_{x,i}^{k+1}$ to its neighbors;
		\State 3. Update $\hat{\boldsymbol{\alpha}}_i^{k+1} = \hat{\boldsymbol{\alpha}}_i^k + c(|\mathcal{N}_i| \hat{\mathbf{x}}_i^{k+1}- \sum_{j \in \mathcal{N}_i} \hat{\mathbf{x}}_j^{k+1} )$;
		\EndFor
	\end{algorithmic}
\end{algorithm*}
%%%%%%%%%%%%%

{\color{black} The noisy ADMM algorithm can be derived by replacing $\mathbf{x}^{k+1}$ with $\hat{\mathbf{x}}^{k+1}$ in \eqref{admmB} and \eqref{admmC}. We have
	\begin{subequations}\label{admmIterationnoisy}
		\begin{equation}\label{admmnoisyA}
		\nabla f(\mathbf{x}^{k+1}) + \mathbf{A}^T\hat{\boldsymbol{\lambda}}^k + c\mathbf{A}^T(\mathbf{A}\mathbf{x}^{k+1} + \mathbf{B}\hat{\mathbf{z}}^k) = \mathbf{0},
		\end{equation}
		\begin{equation}\label{admmnoisyB}
		\mathbf{B}^T\boldsymbol{\hat{\lambda}}^k + c\mathbf{B}^T(\mathbf{A}\hat{\mathbf{x}}^{k+1} + \mathbf{B}\hat{\mathbf{z}}^{k+1}) = \mathbf{0},
		\end{equation}
		\begin{equation}\label{admmnoisyC}
		\hat{\boldsymbol{\lambda}}^{k+1} - \hat{\boldsymbol{\lambda}}^k - c(\mathbf{A}\hat{\mathbf{x}}^{k+1} + \mathbf{B}\hat{\mathbf{z}}^{k+1}) = \mathbf{0}.
		\end{equation}
	\end{subequations}
	Note that $\hat{\mathbf{x}}^{k+1}$, $\hat{\mathbf{z}}^{k+1}$ and $\hat{\boldsymbol{\lambda}}^{k+1}$ represent the variables affected by node noise $\mathbf{e}_x^{k+1}$. Considering $\hat{\boldsymbol{\lambda}} = [\hat{\boldsymbol{\beta}};\hat{\boldsymbol{\nu}}]$ with $\hat{\boldsymbol{\beta}},\hat{\boldsymbol{\nu}} \in \mathbb{R}^{2nE}$, and letting $\mathbf{z}^{k+1} = \frac{1}{2}\mathbf{M}_+^T\mathbf{x}^{k + 1}$ and $\boldsymbol{\beta}^{k+1} = \hat{\boldsymbol{\beta}}^k + \frac{c}{2}\mathbf{M}_-^T\mathbf{x}^{k+1}$, with some mathematical manipulations similar to those done in the derivation of \eqref{midForm}, we have
	\begin{subequations}\label{noisyIter}
		\begin{equation}\label{noisyIterA}
		\nabla f(\mathbf{x}^{k+1}) + \mathbf{M}_-\boldsymbol{\beta}^{k+1} - c\mathbf{M}_+(\hat{\mathbf{z}}^k - \mathbf{z}^{k+1}) = 0,
		\end{equation}
		\begin{equation}\label{noisyIterB}
		\boldsymbol{\beta}^{k+1} = \hat{\boldsymbol{\beta}}^k + \frac{c}{2}\mathbf{M}_-^T\mathbf{x}^{k+1},
		\end{equation}
		\begin{equation}\label{noisyIterC}
		\mathbf{z}^{k+1} = \frac{1}{2}\mathbf{M}_+^T\mathbf{x}^{k + 1},
		\end{equation}
		\begin{equation}\label{noisyIterB2}
		\hat{\boldsymbol{\beta}}^{k+1} = \boldsymbol{\beta}^{k+1} + \frac{c}{2}\mathbf{M}_-^T\mathbf{e}_{x}^{k+1},
		\end{equation}
		\begin{equation}\label{noisyIterC2}
		\hat{\mathbf{z}}^{k+1} = \mathbf{z}^{k+1} + \frac{1}{2}\mathbf{M}_+^T\mathbf{e}_x^{k + 1},
		\end{equation}
	\end{subequations}	
	which can be considered as the noisy version of iteration in \eqref{midForm}. With some algebraic manipulations, this iteration results in the distributed noisy ADMM algorithm outlined in Algorithm \ref{Alg2}.}

	In the following theorem we analyze the convergence behavior of $d^k$ where
	\begin{equation}\label{eq:dDef}
	d^k = \|\hat{\mathbf{u}}^k - \mathbf{u}^*\|_G^2,
	\end{equation}
	and
	\begin{equation}\label{eq:uHat}
	\hat{\mathbf{u}}^k =
	\begin{pmatrix}
	\hat{\mathbf{z}}^k\\
	\hat{\boldsymbol{\beta}}^k
	\end{pmatrix},
	\end{equation}
	and $\mathbf{u}^*$ and $\mathbf{G}$ are defined in (\ref{uGDef}). From \eqref{noisyIter} it can be seen that $d^k$ is a random variable, whose randomness is caused by error sequence $\{\mathbf{e}_x^1,\dots,\mathbf{e}_x^k\}$. In Theorem \ref{linearConvTheo}, we obtain an upper bound on $\underset{k\rightarrow \infty}{\operatorname{lim}}\mathbb{E}[d^{k}]$, which shows that $\underset{k \rightarrow \infty}{\lim}\hspace{1mm}d^k $ is bounded. As we shall demonstrate in proof of this theorem, boundedness of $\underset{k \rightarrow \infty}{\lim}\hspace{1mm}d^k $ dictates that $\underset{k \rightarrow \infty}{\lim}\hspace{1mm}\|\mathbf{x}^k - \mathbf{x}^*\|_2^2$ is finite. We provide an upper bound on the mean squared steady state error $\underset{k\rightarrow \infty}{\operatorname{lim}} \mathbb{E}[\|\mathbf{x}^{k+1} - \mathbf{x}^*\|_2^2]$ of the algorithm.  The proof of the theorem requires the KKT conditions for (\ref{matrixForm}) to hold, i.e.,
	\begin{subequations}\label{eq:KKTCond}
		\begin{equation}\label{eq:KKT1}
		\nabla f(\mathbf{x}^*) + \mathbf{M}_-\boldsymbol{\beta}^* = 0,
		\end{equation}
		\begin{equation}\label{eq:KKT2}
		\mathbf{M}_-^T\mathbf{x}^* = 0,
		\end{equation}
		\begin{equation}\label{eq:KKT3}
		\frac{1}{2}\mathbf{M}_+^T\mathbf{x}^* - \mathbf{z}^* = 0.
		\end{equation}
	\end{subequations}
	%%%% Theorem2

	\begin{theo}\label{linearConvTheo}
Consider the optimization problem in (\ref{matrixForm}) and the iterative solution in (\ref{noisyIter}) for the optimal points $\mathbf{x}^*$ and $\mathbf{z}^*$. The elements of noise vector $\mathbf{e}_x^{k + 1}$ are zero mean i.i.d random variables whose variance are $\sigma_n^2$. Assume that local objective functions satisfy Assumption \ref{assum1}, and the initial value $\boldsymbol{\beta}^0$ lies in the column space of $\mathbf{M}_-^T$. We have
			%\begin{equation}\label{eq:dKConv}
			%\underset{k\rightarrow \infty}{\operatorname{lim}}\mathbb{E}[d^{k}] \leq \frac{1+\delta}{\delta}\zeta,
			%\end{equation}
			%where $\delta$ and $d^{k}$ are defined in (\ref{delta}) and (\ref{eq:dDef}), respectively, and			
			%In addition, we have
			{\color{black}
			\begin{equation}\label{eq:xUpperBound}
			\underset{k\rightarrow \infty}{\operatorname{lim}} \mathbb{E}_{\mathcal{F}^k}[\|\mathbf{x}^{k} - \mathbf{x}^*\|_2^2] \leq \frac{(4 + 3\delta)}{\delta(m_f + 2c\lambda_{min}(\mathbf{W}))}2cnE\sigma_n^2,
			\end{equation}}
			where $\mathcal{F}^k = \{\mathbf{e}_x^1,\dots,\mathbf{e}_x^k\}$ is the history of observed error vectors up to iteration $k$,
			%\end{equation}
			 and $\lambda_{min}(\mathbf{W})$ represents the minimum eigenvalue of $\mathbf{W}$ defined in \eqref{eq:wDef}.
		\end{theo}
			
		%%%%Proof	
	\begin{IEEEproof}
See Appendix \ref{ap:proof2}.
	\end{IEEEproof}
\begin{remark}
  In the noiseless scenario, we have $\sigma_n^2 = 0$, and Theorem \ref{linearConvTheo} indicates that the upper bound on mean squared steady state error is zero. This implies that $\textbf{x}^*$ converges to the optimal point of problem \eqref{matrixForm}.
\end{remark}
\begin{remark}
  The upper bound on the mean squared steady state error in \eqref{eq:xUpperBound} depends on the network topology through $\delta$, $E$ and the minimum eigenvalue of matrix $\mathbf{W}$. There is no explicit mathematical relationship between the network topology and these parameters. We investigate this through some numerical experiments in Section \ref{sec:numerical}.
\end{remark}
		%%%%%%%%%%%%%%%%%%%%%%%%%%%%
		%%%%%%%%%%%%%%%%%%%%%%
		Theorem \ref{linearConvTheo} provides an upper bound on the mean squared steady state error of the estimate produced by Algorithm \ref{Alg2} under Assumption \ref{assum1}. We obtain a lower bound on the mean squared steady state error in Theorem
		\ref{theo:lowerBound}.
		%%%%%%%%%%%%%%%%%%%%%%%% Theorem 3 : lower bound
		
		\begin{theo}\label{theo:lowerBound}
				Consider the optimization problem (\ref{matrixForm}) and iterative solution (\ref{noisyIter}) for optimal points $\mathbf{x}^*$ and $\mathbf{z}^*$. The elements of noise vector $\mathbf{e}_x^{k + 1}$ are zero mean i.i.d random variables whose variance are $\sigma_n^2$. Assume that the local objective functions satisfy Assumption \ref{assum1}. We have
				\begin{equation}\label{lowerBound}
				\mathbb{E}_{\mathcal{F}^{k+1}}\big[\|\mathbf{x}^{k+1} - \mathbf{x}^*\|_2^2\big] \geq \frac{8nEc^2\sigma_n^2}{\big(M_f + 2c\sigma_{max}(\mathbf{W})\big)^2}.
				\end{equation}
			\end{theo}
			%%%%%%%%%%%%%%%%%%%%%%%%%%%%%%%%%
			
			\begin{IEEEproof}
				See Appendix \ref{ap:proof3}.
			\end{IEEEproof}
			%%%%%%%%%	Theorem4 : upper bound non-strongly convex
		%%%%%%%%%%%%%%%%%%%%%%%%%%%
		%%%%%%%%%%%%%%%%%%%		
		Theorem \ref{linearConvTheo} provides an upper bound on the mean squared steady state error of the estimate produced by Algorithm \ref{Alg2} under Assumption \ref{assum1}. In the following theorem, we will show that the steady state error of the algorithm is bounded under a milder condition. To enable the proof we need Proposition \ref{prep:boydConv} derived from the ADMM proof of convergence provided in \cite{boyd2011distributed}, Proposition \ref{prep:ramark3} and Corollary \ref{cor:boundTermDec}.
			%%%%%%%%%%%%%
		\begin{prep}\label{prep:boydConv}
			Consider the optimization problem in the form of problem in \eqref{matrixForm}, and iteration in \eqref{admmIteration} to solve it, for the general $f(\mathbf{x})$ and $g(\mathbf{z})$ functions, and $\mathbf{A}$ and $\mathbf{B}$ matrices.
Assume that $f(\mathbf{x})$ and $g(\mathbf{z})$ are convex, proper and closed functions, and the corresponding Lagrangian function has a saddle point $(\mathbf{x}^*,\mathbf{z}^*,\boldsymbol{\lambda}^*)$. Let
		\begin{equation}\label{eq:layapun}
		V^k \triangleq \frac{1}{c}\|\boldsymbol{\lambda}^k - \boldsymbol{\lambda}^*\|_2^2 + c\|\mathbf{B}(\mathbf{z}^{k} - \mathbf{z}^*)\|_2^2.
		\end{equation}
		 We have
		\begin{equation}\label{eq:admmConvLy}
		V^{k+1} \leq V^k - c\|\mathbf{A}\mathbf{x}^{k+1} + \mathbf{B}\mathbf{z}^{k+1}\|_2^2  - c\|\mathbf{B}(\mathbf{z}^{k+1} - \mathbf{z}^k)\|_2^2,
		\end{equation}
and
\begin{equation}\label{eq:rZLimit}
		\underset{k \rightarrow \infty}{\lim}\hspace{1mm}\mathbf{A}\mathbf{x}^{k} + \mathbf{B}\mathbf{z}^{k}  = \mathbf{0}, \hspace{3mm} \underset{k \rightarrow \infty}{\lim}\hspace{1mm}\mathbf{B}(\mathbf{z}^{k+1} - \mathbf{z}^k) = \mathbf{0}.
		\end{equation}
		\end{prep}
		\begin{IEEEproof}
			See Appendix of \cite{boyd2011distributed}.
		\end{IEEEproof}
		%%%%%%%%%%%%%
		\begin{prep}\label{prep:ramark3}
Consider the optimization problem in \eqref{matrixForm}, definition of matrices $\mathbf{A}$ and $\mathbf{B}$, and the corresponding iterative solution in \eqref{midForm}. If the local objective functions are convex, proper and closed, the iterative solution converges to an optimal point of the problem.
        \end{prep}
\begin{IEEEproof}
%%%%%%%%%%%
%\begin{remark}\label{rem:ourLyapFunct}
See Appendix \ref{ap:prop2}.
\end{IEEEproof}
		%\end{remark}
		%%%%%%%%%
		\begin{corol}\label{cor:boundTermDec}
 We have
\begin{equation}\label{eq:linearConvNonStr}
		\begin{split}
		&\frac{1}{c}\|\boldsymbol{\beta}^{k+1} - \boldsymbol{\beta}^*\|_2^2 + c\|\mathbf{z}^{k+1} - \mathbf{z}^*\|_2^2\\
		&\leq \frac{1}{1 + \delta^{k+1}}\big(\frac{1}{c}\|\boldsymbol{\beta}^{k} - \boldsymbol{\beta}^*\|_2^2 + c\|\mathbf{z}^{k} - \mathbf{z}^*\|_2^2\big),
		\end{split}
		\end{equation}
where $\delta^0 = 0$ and $\delta^{k+1} > 0,\forall k\geq0$.
\end{corol}
\begin{IEEEproof}		
See Appendix \ref{ap:Cor1}.
		\end{IEEEproof}
		%%%%%%%%%%%%%
In the following theorem we show that the steady state error of the noisy ADMM is bounded when local objective functions are closed, proper and convex, and do not necessarily meet more conditions, such as strong convexity.
		\begin{theo}\label{theo:boundNonStr}
		Consider the optimization problem in (\ref{matrixForm}) and the iterative solution in (\ref{noisyIter}) for the optimal points $\mathbf{x}^*$ and $\mathbf{z}^*$. The elements of noise vector $\mathbf{e}_x^{k + 1}$ are zero mean i.i.d random variables whose variance are $\sigma_n^2$, and $\|\mathbf{e}_x^{k}\|_2 \leq e_{max} < \infty$. Assume that the local objective functions are convex, closed and proper. We almost surely have
		\begin{equation}\label{eq:uBoundNonStr}
			\underset{k \rightarrow \infty}{\lim} \hspace{1mm}\|\mathbf{x}^k - \mathbf{x}^*\|_2^2 < \infty,
		\end{equation}
		 i.e., the steady state error of the Algorithm 2 is bounded.
		\end{theo}
		\begin{IEEEproof}
See Appendix \ref{ap:Theorem4}
		\end{IEEEproof}			
	
            %%%%%%%%%%		
			\section{Numerical Evaluation}\label{sec:numerical}
			In this section, we use Algorithm \ref{Alg2} to solve the following optimization problem
				\begin{equation}\label{LSE}
				\min_{\tilde{\mathbf{x}}}\sum_{i=1}^{N}\frac{1}{2}\|\mathbf{y}_i - \mathbf{M}_i\tilde{\mathbf{x}}\|_2^2.
				\end{equation}
			$N$ agents estimate variable $\tilde{\mathbf{x}}$ using noisy linear observations $\mathbf{y}_i = \mathbf{M}_i\tilde{\mathbf{x}} + \mathbf{n}_i$, cooperatively. The variables communicated between neighbors are subject to additive node error as in (\ref{noisy}). In these experiments, the elements of error vector are i.i.d and follow uniform distribution $\mathcal{U}(-\epsilon,\epsilon)$.  We consider network edge density as $r \triangleq \frac{E}{E_c}$, where $E_c$ is the number of edges in a corresponding complete graph. Consider $\mathbf{x}_{k}^D$ and $\mathbf{x}^C$ as respectively the distributed and centralized estimates of $\mathbf{x}$ (at iteration $k$). Our performance metrics are the relative error $\sqrt{\mathbb{E}[\mathcal{E}^{DC}_{k}]}$, where $\mathcal{E}^{DC}_{k} = \frac{\|\mathbf{x}^D_{k} - \mathbf{x}^C_{}\|_2^2}{\|\mathbf{x}^C\|_2^2}$, and steady state error $\underset{k\rightarrow \infty}{\operatorname{lim}} \sqrt{\mathbb{E}[\mathcal{E}^{DC}_{k}]}$.
			
			In the first set of experiments, the number of agents $N$ is $20$, and $\tilde{\mathbf{x}} \in \mathbb{R}^{20}$ (i.e., $n=20$) and its elements are i.i.d from $\mathcal{N}(0,1)$. Measurement matrices $\mathbf{M}_i \in \mathbb{R}^{5 \times 20}$, which means that local objective functions are not strongly convex. The observation noise is $\mathbf{n}_i \sim \mathcal{N}(\mathbf{0},\sigma^2\mathbf{I}_5)$ and $\sigma^2 = 0.1$. Figure \ref{fig:normNonStr} presents the relative error of the algorithm as a function of iteration index $k$ for different values of $\epsilon$. In this experiment $c = 1$ and $r = 0.5$. As evident the algorithm is convergent, which confirms theoretical result form Theorem \ref{theo:boundNonStr}.			
			
			\begin{figure}[!t]
				%\begin{minipage}[b]{1.0\linewidth}
				\centering
				\centerline{\includegraphics[width=8cm]{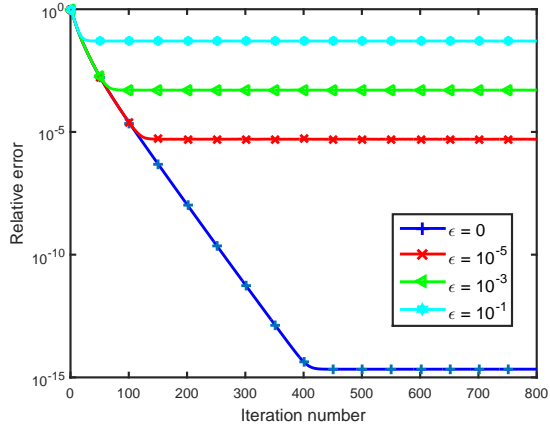}}
				%\vspace{2.0cm}
				%\centerline{(a)}\medskip
				%\end{minipage}
				\caption{Performance of
Algorithm \ref{Alg2} when local objective functions are not strongly convex: relative error vs. iteration number over randomly generated networks for $N=20$, $c =  1$, $r = 0.5$}
				\label{fig:normNonStr}
			\end{figure}	
					%\begin{figure}[!t]
				%\begin{minipage}[b]{1.0\linewidth}
			%	\centering
			%	\centerline{\includegraphics[width=9.5cm]{nonStrComp}}
				%\vspace{2.0cm}
				%\centerline{(a)}\medskip
				%\end{minipage}
			%	\caption{Performance of ADMM and DGD algorithms in presence of node error when measurement matrices $\mathbf{M}_i$ are fat, obtained with randomly generated networks for $\epsilon = 10^{-5}$, $N=20$, and $r = 0.5$.}
			%	\label{fig:compNonStr}
			%\end{figure}				
			In the second set of experiments the local objective functions satisfy Assumption \ref{assum1}, and $\tilde{\mathbf{x}} \in \mathbb{R}^3$ (i.e., $n=3$) and its elements are i.i.d from $\mathcal{N}(0,1)$. We first generate measurement matrices $\mathbf{M}_i \in \mathbb{R}^{3 \times 3}$, whose elements are i.i.d from $\mathcal{N}(0,1)$. Then, scale their singular values to the range $[\sqrt{m_f},\sqrt{M_f}]$ and rebuild $\mathbf{M}_i$. The observation noise is $\mathbf{n}_i \sim \mathcal{N}(\mathbf{0},\sigma^2\mathbf{I}_3)$ and $\sigma^2 = 0.1$. In the following experiments, unless otherwise stated, the parameter $c$ is set to $c^*$ in (\ref{eq:cOpt}).				
			
			The upper bound in Theorem \ref{linearConvTheo} depends on $\delta$ whose theoretical value is presented in (\ref{delta}). In numerical experiments, we can compute this parameter by estimating convergence rate of the algorithm experimentally. Define $\rho_e^k$ as the estimated convergence rate of the algorithm at iteration $k$
				\begin{equation}\label{eq:rhoP}
			\rho_e^k \triangleq \frac{\|\mathbf{u}^{k} - \mathbf{u}^*\|_G^2}{\|\mathbf{u}^{k-1} - \mathbf{u}^*\|_G^2}.
			\end{equation}
				The geometric average rate of convergence of the algorithm is
			\begin{equation}\label{eq:rhoAver}
			\bar{\rho}_e = \big(\prod_{k=1}^{K} \rho_e^k\big)^{\frac{1}{K}}.
			\end{equation}
			Based on (\ref{delta}) and (\ref{eq:rhoAver}), we have
			\begin{equation}\label{eq:deltaP}
			\delta_e = \frac{1}{\bar{\rho}_e} -1.
			\end{equation}	
					
	In the sequel, the upper bounds in Theorem \ref{linearConvTheo} which are computed based on different $\delta$ parameters presented in (\ref{delta}) and (\ref{eq:deltaP}) are referred to as theoretical upper bound, $\text{UB}_t$, and experimental upper bound, $\text{UB}_e$, respectively. The lower bound in Theorem \ref{theo:lowerBound} is labeled as $\text{LB}$.
				
			Figure \ref{fig:errorNorm} depicts the relative error of  Algorithm \ref{Alg2} as a function of iteration number $k$. In this experiment, the nodes communicate over a randomly generated connected network of $N = 20$ nodes with edge density $0.5$, and $\epsilon = 10^{-4}$. The performance of the algorithm in error free scenario is also presented for comparison. As seen, $\text{UB}_e$ is a tighter bound than $\text{UB}_t$, indicating that experimentally computed convergence rate is more accurate.
		
A distributed subgradient method is proposed in \cite{nedic2008distributed}, which is convergent in presence of quantization error. Figure \ref{fig:DGDComp} presents the performance of this algorithm, labeled as DGD and Algorithm \ref{Alg2} (ADMM), for comparison. In this experiment, we have $N = 20$ and $r = 0.5$. This figure shows that the ADMM-based algorithm outperforms DGD algorithm.
			
%Figure \ref{fig:DGDComp} presents the performance of DGD algorithm  proposed in \cite{nedic2008distributed}, and Algorithm \ref{Alg2} (ADMM), for comparison. In this experiment, we have $N = 20$ and $r = 0.5$. This figure shows that similar to first set of experiments, the ADMM-based algorithm outperforms DGD algorithm.				
			Figure \ref{fig:epsilon} shows the effect of $\epsilon$ on steady state error of Algorithm \ref{Alg2}. In the current experiment set up with the least squares as the objective function of the optimization problem, one can show that indeed the steady state error of the algorithm is a linear function of the node error. This holds true not only in the performance of the algorithm (as reflected in simulations), but also in the presented lower and upper bounds.
				\begin{figure}[!t]				
				\centering
				\centerline{\includegraphics[width=8cm]{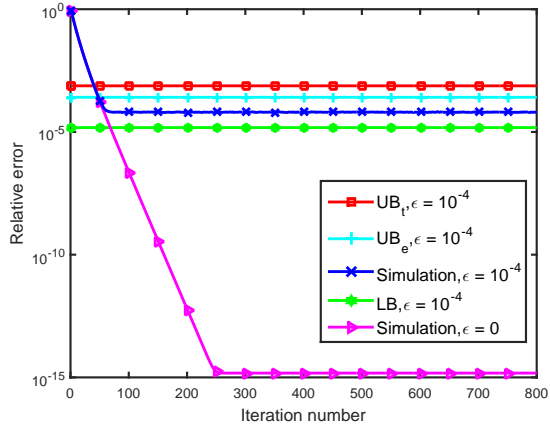}}				
				\caption{Performance of Algorithm \ref{Alg2} when local objective functions are strongly convex and their gradients are Lipschitz continuous : relative error vs. iteration number over randomly generated networks for $N=20$, $M_f = 10$, $m_f = 1$, $r = 0.5$}
					\label{fig:errorNorm}
				\end{figure}
				
				\begin{figure}[!t]				
				\centering
				\centerline{\includegraphics[width=8cm]{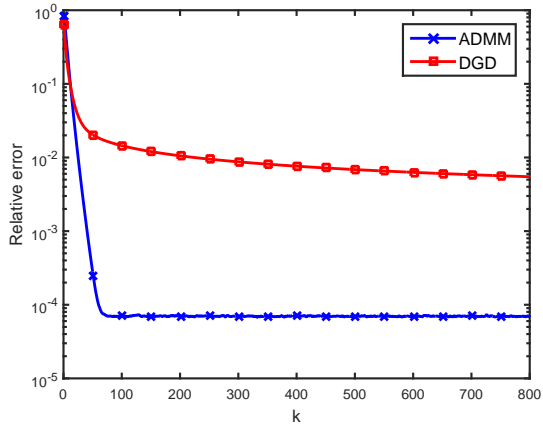}}				
				\caption{Performance of Algorithm \ref{Alg2} (ADMM) and DGD algorithm \cite{nedic2008distributed}, obtained with randomly generated networks for $\epsilon = 10^{-4}$, $N=20$, $M_f = 10$, $m_f = 1$, and $r = 0.5$.}
				\label{fig:DGDComp}
				\end{figure}
				
				\begin{figure}[!t]					
					\centering
					\centerline{\includegraphics[width=8cm]{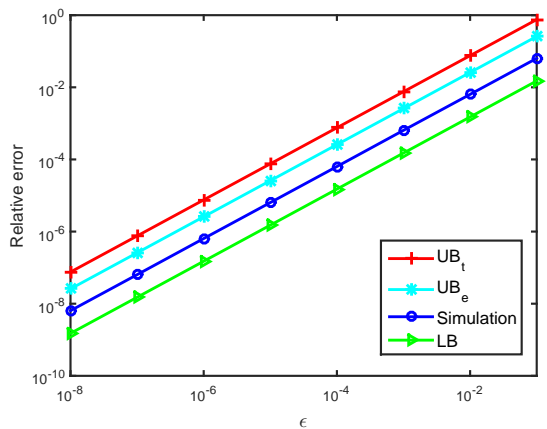}}					
					\caption{Steady state error vs. $\epsilon$, obtained with randomly generated networks, $N=20$, $M_f = 10$, $m_f = 1$, $r = 0.5$}
					\label{fig:epsilon}
				\end{figure}
				
				Figure \ref{fig:ed} presents the influence of network edge density on the steady state error of the algorithm. In this experiment networks of $N=20$ and $N=100$ nodes are generated randomly. One sees that increase in network edge density leads to little decrease in steady state error. The trends of simulation results and lower and upper bounds are consistent.
				
				\begin{figure}[!t]					
					\centering
					\centerline{\includegraphics[width=8cm]{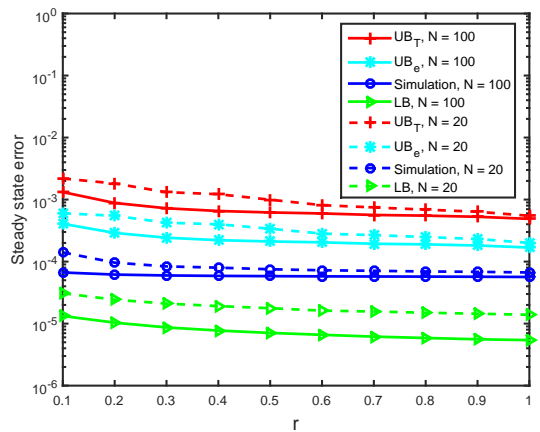}}					
					\caption{Steady state error vs. edge density as a function of number of nodes $N$, obtained with randomly generated networks. $\epsilon = 10^{-4}$, $M_f = 10$, $m_f = 1$.}
					\label{fig:ed}
				\end{figure}
				
				We study the performance of the noisy ADMM algorithm over regular networks as an especial case. Figure \ref{fig:degEff} depicts the effect of node degree on the steady state error of the algorithm for networks of $N = 100$ nodes. As evident, the slop of the curve is a decreasing function of node degree, and varying node degree in the range of $[30\hspace{0.1cm} 99]$ does not improve steady state error significantly. The upper bounds which are functions of $E$ and $\delta$, predict this behavior properly. The lower bound is just a function of $E$, and  does not predict the slop of steady state error curve.
				\begin{figure}[!t]
					%\begin{minipage}[b]{1.0\linewidth}
						\centering
						\centerline{\includegraphics[width=8cm]{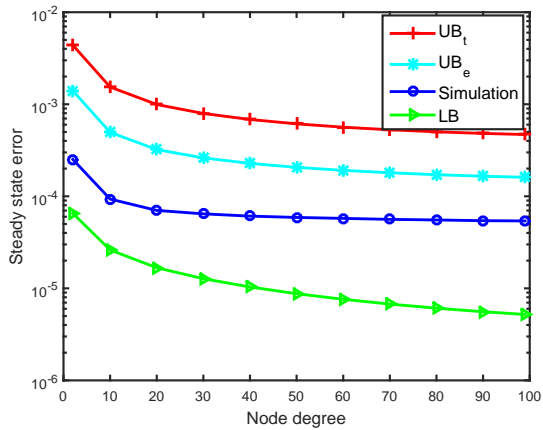}}
						%\vspace{2.0cm}
					%	\centerline{(a)}\medskip
					%\end{minipage}
					%\begin{minipage}[b]{1.0\linewidth}
					%	\centering
					%	\centerline{\includegraphics[width=9.5cm]{DegEffect20_2}}
						%\vspace{2.0cm}
					%	\centerline{(b)}\medskip
					%\end{minipage}
					\caption{Steady state error vs. node degree of regular network, $\epsilon = 10^{-4}$, $M_f = 10$, $m_f = 1$, $N = 100$.}
					\label{fig:degEff}
				\end{figure}
				
				We consider tree network as another especial case. Figure \ref{fig:BF} demonstrates the influence of branch factor parameter of the tree network on
				steady state error of the algorithm. It can be seen that experimental upper bound predicts the performance of the  algorithm better than the other bounds, and proposed theoretical upper bound is an acceptable indicator of the algorithm behavior. From the last three sets of experiments, we can conclude that the upper bounds can predict effects of network topology on the steady state error more accurately.
				\begin{figure}[!t]
				%	\begin{minipage}[b]{1.0\linewidth}
						\centering
						\centerline{\includegraphics[width=8cm]{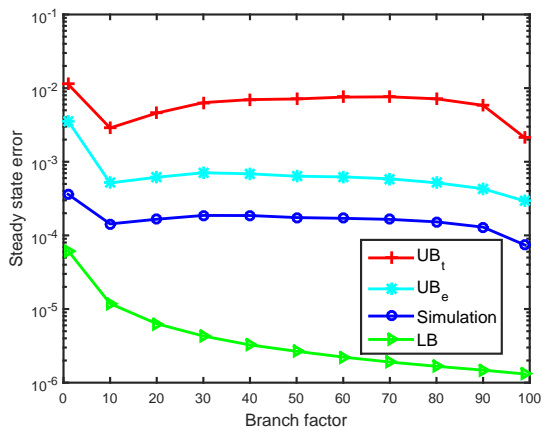}}
						%\vspace{2.0cm}
					%	\centerline{(a)}\medskip
					%\end{minipage}
				%	\begin{minipage}[b]{1.0\linewidth}
					%	\centering
					%	\centerline{\includegraphics[width=9.5cm]{BFEffect20_2}}
						%\vspace{2.0cm}
					%	\centerline{(b)}\medskip
					%\end{minipage}
					\caption{Steady state error vs. branch factor of tree network, $\epsilon = 10^{-4}$, $M_f = 10$, $m_f = 1$, $N = 100$}
						\label{fig:BF}
				\end{figure}		
				%We consider tree network as another especial case. Figure \ref{fig:BF} demonstrates the influence of branch factor parameter of the tree network on steady state error of the algorithm. It can be seen that experimental upper bound predicts the performance of the  algorithm better than the other bounds, and proposed theoretical upper bound is an acceptable indicator of the algorithm behavior. From the last three sets of experiments, we can conclude that the upper bounds can predict effects of network topology on the steady state error more accurately.
				%\begin{figure}[!t]
					%\begin{minipage}[b]{1.0\linewidth}
				%		\centering
				%		\centerline{\includegraphics[width=8cm]{BFEffect100_2}}
						%\vspace{2.0cm}
						%\centerline{(a)}\medskip
					%\end{minipage}
					%\begin{minipage}[b]{1.0\linewidth}
						%\centering
						%\centerline{\includegraphics[width=9.5cm]{BFEffect20_2}}
						%\vspace{2.0cm}
						%\centerline{(b)}\medskip
					%\end{minipage}
					%\caption{Steady state error vs. branch factor of tree network, $\epsilon = 10^{-4}$, $M_f = 10$, $m_f = 1$, $N = 100$.}
					%\label{fig:BF}
				%\end{figure}		
				
Another parameter which affects the performance of the algorithm is $c$. Figure \ref{fig:cParam} shows the effect of $c$ parameter on the steady state error over the randomly generated networks. The connectivity ratio in $N = 20$ and $ N =100$  scenarios are $0.5$ and $0.2$, respectively. One sees that, a lower $c$ results in a lower steady state error which is also predicted by the proposed lower and experimental upper bounds. The mismatch between theoretical upper bound and simulation shows that the effect of $c$ on the convergence rate of the algorithm is not predicted properly by (\ref{eq:rho}).
				
		{\color{black} As discussed, the convergence rate and steady state error of the algorithm output depends on parameter $c$. The value of the parameter $c$ is to be selected for minimized steady state error and maximized convergence rate. According to our observations, the performance of the noisy ADMM algorithm in early iterations is similar to the noiseless case. A smaller value of $c$ results in a smaller steady state error, but it does not necessarily lead to a high convergence rate. A reasonable tuning policy is to set parameter $c$ to accelerate the convergence in early iterations, and then to reduce the parameter to aim for a small steady state error.} Figure 9 depicts the performance of this tuning method. In this experiment, parameter $c$ is set to $c^*$ provided in (14) in the first $200$ iterations, and then it is reduced by a factor of $0.01$. The performance of the algorithm with fixed parameter $c$ set to $0.01c^*$ is also presented for comparison. The results confirm faster convergence rate and improved steady state error when the proposed tuning policy is in effect.
				
				\begin{figure}[!t]					
					\centering
					\centerline{\includegraphics[width=8cm]{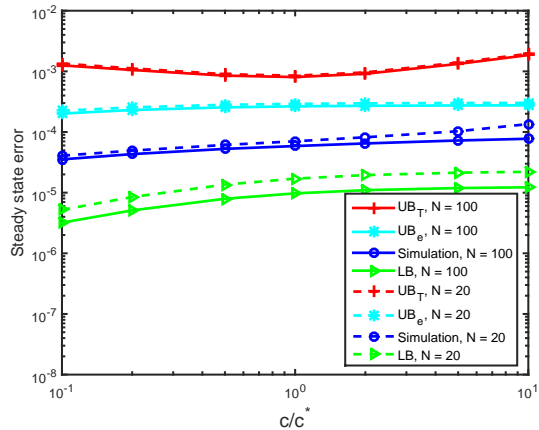}}					
					\caption{Steady state error vs. $c$ parameter as a function of number of nodes $N$, obtained with randomly generated networks. $\epsilon = 10^{-4}$, $M_f = 10$, $m_f = 1$.}
					\label{fig:cParam}
				\end{figure}
				
				\begin{figure}[!t]					
					\centering
					\centerline{\includegraphics[width=8cm]{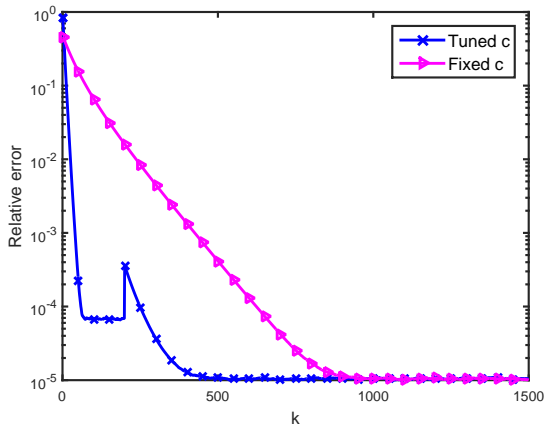}}					
					\caption{Comparison of performance of tuned and fixed $c$ algorithms, $\epsilon =10^{-4}$, $M_f = 10$, $m_f = 1$,$N = 20$, $r = 0.5$}
					\label{fig:cVar}
				\end{figure}
				
				{\color{black}We also examined the effects of $M_f$, Lipschitz constant,  and $m_f$, strong convexity module, on the steady state error of the algorithm (not reported here). We observed that the influence of these parameters is accurately predicted by the proposed lower and experimental upper bounds. The theoretical upper bound, however, does not reflect the effects of these parameters properly.}
				%\begin{figure}[!t]					
				%	\centering
				%	\centerline{\includegraphics[width=9.5cm]{MMMFFEffect_2}}					
				%	\caption{Steady state error vs. $M_f$ parameter as a function of number of nodes $N$, obtained with randomly generated networks. $\epsilon = 10^{-4}$, $m_f = 1$.}
					%\label{fig:Mf}
				%\end{figure}
				
				%\begin{figure}[!t]					
				%	\centering
				%	\centerline{\includegraphics[width=9.5cm]{mmffEffect_2}}					
				%	\caption{Steady state error vs. $m_f$ parameter as a function of number of nodes $N$, obtained with randomly generated networks. $\epsilon = 10^{-4}$, $M_f = 10$.}
				%	\label{fig:mff}
				%\end{figure}
			
			\section{Conclusions}\label{sec:concl}
 We analyzed the effect of additive node error on the performance of distributed ADMM algorithm for consensus optimization over a connected network. Analytical upper and lower bounds were provided on the mean squared steady state error of the algorithm, in the case that local objective functions are strongly convex and have Lipschitz continuous gradients. The analysis quantifies how the provided bounds depend on different factors such as noise variance and network topology. To accelerate the algorithm and also reduce the steady state error, a method was proposed to tune the algorithm parameter. In addition, it was shown that if the local objective functions are proper, closed and convex, for a bounded and random node error, the steady state error of the noisy ADMM algorithm for consensus is bounded. Numerical results validated the theoretical analyses and demonstrated the role of different system and network parameters. Analysis of the algorithm convergence behavior over the networks whose links suffer from additive noise can be considered as an important next research steps.
%%%%%%%%%%%
\appendices
\section{Proof of Theorem 2}\label{ap:proof2}
	We first derive an upper bound on $\underset{k\rightarrow \infty}{\operatorname{lim}}\mathbb{E}[d^{k}]$. We have
	\begin{equation}\label{eq:expDk}
	\mathbb{E}_{\mathcal{F}^{k+1}}[d^{k+1}] = \mathbb{E}_{\mathcal{F}^{k}}\big\{\mathbb{E}_{e_x^{k+1}}[d^{k+1}|\mathcal{F}^{k}]\big\},
	\end{equation}
	where 	
	{\color{black}				
		\begin{equation*}%\label{eq:theo1ProofU}
		\begin{split}
		& \mathbb{E}_{e_x^{k+1}}[d^{k+1} | \mathcal{F}^k] = \mathbb{E}_{e_x^{k+1}}\big[\frac{1}{c}\|\hat{\boldsymbol{\beta}}^{k+1} - \boldsymbol{\beta}^*\|_2^2 + c \|\hat{\mathbf{z}}^{k+1} - \mathbf{z}^*\|_2^2|\mathcal{F}^k\big]  \\
		& \stackrel{\text{(a)}}{=} \mathbb{E}_{e_x^{k+1}}\big[\frac{1}{c}\|\boldsymbol{\beta}^{k+1} - \boldsymbol{\beta}^* + \frac{c}{2}\mathbf{M}_-^T\mathbf{e}_x^{k+1}\|_2^2 | \mathcal{F}^k \big]\\
		& \hspace{0.3 cm} + \mathbb{E}_{e_x^{k+1}}\big[c \|\mathbf{z}^{k+1} - \mathbf{z}^*  + \frac{1}{2}\mathbf{M}_+^T\mathbf{e}_x^{k+1}\|_2^2 | \mathcal{F}^k\big]\\		
		\end{split}
		\end{equation*}	
		\begin{equation}\label{eq:theo1ProofU}
		\begin{split}
		& \stackrel{\text{(b)}}{=} \frac{1}{c}\|\boldsymbol{\beta}^{k+1} - \boldsymbol{\beta}^*\|_2^2 + c \|\mathbf{z}^{k+1} - \mathbf{z}^*\|_2^2 + c\mathbb{E}[(\mathbf{e}_x^{k+1})^T\mathbf{W}\mathbf{e}_x^{k+1}]\\
		& = \frac{1}{c}\|\boldsymbol{\beta}^{k+1} - \boldsymbol{\beta}^*\|_2^2 + c \|\mathbf{z}^{k+1} - \mathbf{z}^*\|_2^2 + 2cnE\sigma_n^2\\			
		& \stackrel{\text{(c)}}{\leq} \frac{1}{1 + \delta}\bigg\{\frac{1}{c}\|\hat{\boldsymbol{\beta}}^{k} - \boldsymbol{\beta}^*\|_2^2 + c \|\hat{\mathbf{z}}^{k} - \mathbf{z}^*\|_2^2\bigg\} + 2cnE\sigma_n^2\\
		& = \frac{1}{1 + \delta}d^k + 2cnE\sigma_n^2,
		\end{split}
		\end{equation}	}					
	where (a) follows from (\ref{noisyIterB2}) and (\ref{noisyIterC2}), (b) follows from  the fact that the elements of noise vector $\mathbf{e}_x^{k+1}$ are zero mean i.i.d random variables and (c) follows from  (\ref{convergence}) and (\ref{noisyIterA})- (\ref{noisyIterC}).
	From \eqref{eq:expDk} and \eqref{eq:theo1ProofU}, we have
	{\color{black}
		\begin{equation}\label{eq:meanDK}
		\mathbb{E}_{\mathcal{F}^{k+1}}[d^{k+1}] \leq \frac{1}{1 + \delta}\mathbb{E}_{\mathcal{F}^{k}}[d^{k}] + 2cnE\sigma_n^2.
		\end{equation}}
	Recursively using \eqref{eq:meanDK}, we have
	{\color{black}
		\begin{equation}\label{eq:meanDkInf}
		\mathbb{E}_{\mathcal{F}^{k+1}}[d^{k+1}] \leq \frac{d^0}{(1 + \delta)^{k+1}} + \sum_{ \ell = 0}^{k}\frac{1}{(1+\delta)^\ell}2cnE\sigma_n^2,
		\end{equation}}
	which leads to
	{\color{black}
		\begin{equation}\label{eq:expLimitD}
		\underset{k\rightarrow \infty}{\operatorname{lim}}\mathbb{E}_{\mathcal{F}^{k}}[d^{k}] \leq \frac{1+\delta}{\delta}2cnE\sigma_n^2.			
		\end{equation}}
	
	To prove (\ref{eq:xUpperBound}), we use the update equation for $\mathbf{x}$ in \eqref{noisyIterA}, and its  corresponding KKT condition \eqref{eq:KKT1}. Our goal is to derive a mathematical relationship between $\mathbf{x}^{k+1} - \mathbf{x}^{*}$, $\hat{\boldsymbol{\beta}}^k - \boldsymbol{\beta}^*$ and $\hat{\mathbf{z}}^k - \mathbf{z}^*$.
	To this aim we combine \eqref{noisyIterA} and \eqref{eq:KKT1} as
	\begin{equation}\label{eq:combKKT}
	\nabla f(\mathbf{x}^{k+1}) - \nabla f(\mathbf{x}^{*}) + \mathbf{M}_-(\boldsymbol{\beta}^{k+1} - \boldsymbol{\beta}^*)
	- c\mathbf{M}_+(\hat{\mathbf{z}}^k - \mathbf{z}^{k+1}) = 0.
	\end{equation}
	Replacing (\ref{noisyIterB}) and (\ref{noisyIterC}) in (\ref{eq:combKKT}), we have
	\begin{equation}\label{eq:combKKT12}
	\begin{split}			
	\nabla f(\mathbf{x}^{k+1}) - &\nabla f(\mathbf{x}^{*}) + \frac{c}{2}\mathbf{M}_- \mathbf{M}_-^T\mathbf{x^{k+1}}  + \frac{c}{2}\mathbf{M}_+\mathbf{M}_+^T\mathbf{x}^{k+1}\\
	& = -\mathbf{M}_-(\hat{\boldsymbol{\beta}}^{k} - \boldsymbol{\beta}^*) + c\mathbf{M}_+\hat{\mathbf{z}}^k.	
	\end{split}		
	\end{equation}	
	According to \eqref{eq:wDef}, combining \eqref{eq:combKKT12} with  (\ref{eq:KKT2}) and (\ref{eq:KKT3}) leads to
	\begin{equation}\label{eq:combKKT2}
	\begin{split}
	\nabla f(\mathbf{x}^{k+1}) &- \nabla f(\mathbf{x}^{*}) + 2c\mathbf{W}(\mathbf{x}^{k+1} - \mathbf{x}^*)\\
	&= -\mathbf{M}_-(\hat{\boldsymbol{\beta}}^k - \boldsymbol{\beta}^*) + c\mathbf{M}_+(\hat{\mathbf{z}}^k - \mathbf{z}^*).
	\end{split}
	\end{equation}		
	%%%%%%%%%	
	The inner product of $\mathbf{x}^{k+1} - \mathbf{x}^*$ and (\ref{eq:combKKT2}) is
	\begin{equation*}
	\begin{split}
	&\hspace{-1.5cm}\langle \mathbf{x}^{k+1} - \mathbf{x}^*,\nabla f(\mathbf{x}^{k+1}) - \nabla f(\mathbf{x}^{*})\rangle \\
	&\hspace{-1.2cm}+ 2c \langle \mathbf{x}^{k+1} - \mathbf{x}^*,\mathbf{W}(\mathbf{x}^{k+1} - \mathbf{x}^*)\rangle\\				
	&\hspace{-1.5cm}= -\langle \mathbf{M}_-^T( \mathbf{x}^{k+1} - \mathbf{x}^*), \hat{\boldsymbol{\beta}}^k - \boldsymbol{\beta}^*\rangle\\
	&\hspace{-1.2cm}+ c\langle \mathbf{M}_+^T( \mathbf{x}^{k+1} - \mathbf{x}^*), \hat{\mathbf{z}}^k - \mathbf{z}^*\rangle\\
	\end{split}
	\end{equation*}
	\begin{equation}\label{eq:innerProd}
	\begin{split}	
	&\hspace{-1.5cm}\stackrel{\text{(a)}}{=}-\frac{2}{c}\langle \hat{\boldsymbol{\beta}}^k - \boldsymbol{\beta}^*, (\boldsymbol{\beta}^{k+1} - \boldsymbol{\beta}^*) -(\hat{\boldsymbol{\beta}}^k - \boldsymbol{\beta}^*)\rangle \\			
	&\hspace{0.4cm}+ 2c\langle \hat{\mathbf{z}}^k - \mathbf{z}^*, \mathbf{z}^{k+1} - \mathbf{z}^*\rangle,
	\end{split}
	\end{equation}			
	where $(a)$ follows from  (\ref{noisyIterB}) and (\ref{noisyIterC}). Since $\mathbf{W}$ is a positive definite matrix, and  based on strong convexity of $f(\mathbf{x})$, the lower bound on the LHS of \eqref{eq:innerProd} is
	\begin{equation}\label{eq:lBLHS}
	\big(m_f + 2c\lambda_{min}(\mathbf{W})\big)\|\mathbf{x}^{k+1} - \mathbf{x}^*\|_2^2 .
	\end{equation}
	Replacing \eqref{eq:lBLHS} in \eqref{eq:innerProd} leads to		
	\begin{equation*}%\label{eq:upperB1}
	\begin{split}
	&\hspace{0cm}\big(m_f + 2c\lambda_{min}(\mathbf{W})\big)\|\mathbf{x}^{k+1} - \mathbf{x}^*\|_2^2 \\
	&\hspace{0cm}\leq -\frac{2}{c}\langle \hat{\boldsymbol{\beta}}^k - \boldsymbol{\beta}^*, (\boldsymbol{\beta}^{k+1} - \boldsymbol{\beta}^*) -(\hat{\boldsymbol{\beta}}^k - \boldsymbol{\beta}^*)\rangle \\            		
	&\hspace{0.3cm}+ 2c\langle \hat{\mathbf{z}}^k - \mathbf{z}^*, \mathbf{z}^{k+1} - \mathbf{z}^*\rangle\\			
	&\leq \frac{1}{c}\|\boldsymbol{\beta}^{k+1} - \boldsymbol{\beta}^*\|_2^2 +\frac{3}{c}\|\hat{\boldsymbol{\beta}}^k - \boldsymbol{\beta}^*\|_2^2+ c\|\hat{\mathbf{z}}^k - \mathbf{z}^*\|_2^2 \\			
	&\hspace{0.3cm} + c\| \mathbf{z}^{k+1} - \mathbf{z}^*\|_2^2\\
	\end{split}
	\end{equation*}	
	\begin{equation}\label{eq:upperB1}
	\begin{split}
	&\hspace{-3.5cm}\stackrel{(a)}{\leq} \big(3 + \frac{1}{1 + \delta}\big)\|\hat{\mathbf{u}}^k - \mathbf{u}^*\|_G^2\\
	&\hspace{-3.5cm}\stackrel{(b)}{=}\big(3 + \frac{1}{1 + \delta}\big)d^k,
	\end{split}
	\end{equation}			
	where $(a)$ and $(b)$ follow from \eqref{uGDef} and (\ref{convergence}), and \eqref{eq:dDef}, respectively. Combining (\ref{eq:upperB1}) and (\ref{eq:expLimitD}), we have
	{\color{black}
		\begin{equation}\label{eq:xUBound}
		\underset{k\rightarrow \infty}{\operatorname{lim}} \mathbb{E}_{\mathcal{F}^{k+1}}[\|\mathbf{x}^{k+1} - \mathbf{x}^*\|_2^2] \leq \frac{4 + 3\delta}{\delta(m_f + 2c\lambda_{min}(\mathbf{W}))}2cnE\sigma_n^2,
		\end{equation}}
	which completes the proof.	
	%%%%%%
	\section{Proof of Theorem 3}\label{ap:proof3}
	Subtracting (\ref{eq:KKT1}) from (\ref{noisyIterA}), we have
	\begin{equation}\label{eq:loBound1}
	\nabla f(\mathbf{x}^{k+1}) - \nabla f(\mathbf{x}^{*}) + \mathbf{M}_-(\boldsymbol{\beta}^{k+1} - \boldsymbol{\beta}^*) -c\mathbf{M}_+(\hat{\mathbf{z}}^{k} - \mathbf{z}^{k+1}) = 0.
	\end{equation}
	Replacing (\ref{noisyIterB}) and (\ref{noisyIterC}) in (\ref{eq:loBound1}) yields
	\begin{equation}\label{eq:loBound2}
	\begin{split}
	\nabla f(\mathbf{x}^{k+1})& - \nabla f(\mathbf{x}^{*})  + 2c\mathbf{W}(\mathbf{x}^{k+1} - \mathbf{x}^*)\\
	& =  c\mathbf{M}_+(\hat{\mathbf{z}}^{k} - \mathbf{z}^{*}) - \mathbf{M}_-(\hat{\boldsymbol{\beta}}^{k} - \boldsymbol{\beta}^*).
	\end{split}
	\end{equation}
	Writing $\hat{\mathbf{z}}^{k}$ and $\hat{\boldsymbol{\beta}}^{k}$ in terms of $\mathbf{z}^{k}$, $\boldsymbol{\beta}^{k}$ and $\mathbf{e}_x^k$ in  \eqref{eq:loBound2} leads to \small
	\begin{equation}\label{eq:loBound3}
	\begin{split}
	&\|\nabla f(\mathbf{x}^{k+1}) - \nabla f(\mathbf{x}^{*}) + 2c\mathbf{W}(\mathbf{x}^{k+1} - \mathbf{x}^*)\|_2^2\\
	& = \|c\mathbf{M}_+(\mathbf{z}^{k} - \mathbf{z}^{*}) - \mathbf{M}_-(\boldsymbol{\beta}^{k} - \boldsymbol{\beta}^*)\\ &\hspace{5mm}+\frac{c}{2} (\mathbf{M}_+\mathbf{M}_+^T - \mathbf{M}_-\mathbf{M}_-^T)\mathbf{e}_x^k\|_2^2.
	\end{split}
	\end{equation}
	\normalsize
	The upper bound on the square root of the LHS of (\ref{eq:loBound3}) is
	\begin{equation}\label{eq:loBound4}
	\begin{split}
	&\|\nabla f(\mathbf{x}^{k+1}) - \nabla f(\mathbf{x}^{*}) + 2c\mathbf{W}(\mathbf{x}^{k+1} - \mathbf{x}^*)\|_2 \\
	&\leq \|\nabla f(\mathbf{x}^{k+1}) - \nabla f(\mathbf{x}^{*})\|_2 + \|2c\mathbf{W}(\mathbf{x}^{k+1} - \mathbf{x}^*)\|_2\\
	& \stackrel{\text{(a)}}{\leq }(M_f + 2c\sigma_{max}(\mathbf{W}))\|\mathbf{x}^{k+1} - \mathbf{x}^*\|_2,
	\end{split}
	\end{equation}
	where $(a)$ follows from Lipschitz continuity property of $ \nabla f(\mathbf{x})$. The RHS of (\ref{eq:loBound3}) can be rewritten as follows				
	\small
	%\begin{equation*}
	%\begin{split}
	\begin{equation}\label{eq:loBound44}
	\begin{split}
	&\|c\mathbf{M}_+(\mathbf{z}^{k} - \mathbf{z}^{*}) - \mathbf{M}_-(\boldsymbol{\beta}^{k} - \boldsymbol{\beta}^*) + \frac{c}{2} (\mathbf{M}_+\mathbf{M}_+^T - \mathbf{M}_-\mathbf{M}_-^T)\mathbf{e}_x^k\|_2^2 \\
	%\end{split}
	%\end{equation*}	
	& = 2\langle c\mathbf{M}_+(\mathbf{z}^{k} - \mathbf{z}^{*}) - \mathbf{M}_-(\boldsymbol{\beta}^{k} - \boldsymbol{\beta}^*) , \frac{c}{2} (\mathbf{M}_+\mathbf{M}_+^T - \mathbf{M}_-\mathbf{M}_-^T)\mathbf{e}_x^k\rangle\\			
	& \hspace{3mm}+ \|c\mathbf{M}_+(\mathbf{z}^{k} - \mathbf{z}^{*}) - \mathbf{M}_-(\boldsymbol{\beta}^{k} - \boldsymbol{\beta}^*)\|_2^2\\
	& \hspace{3mm}+ \|\frac{c}{2} (\mathbf{M}_+\mathbf{M}_+^T - \mathbf{M}_-\mathbf{M}_-^T)\mathbf{e}_x^k\|_2^2.				
	\end{split}
	\end{equation}
	\normalsize
	From (\ref{eq:loBound3}), (\ref{eq:loBound4}) and (\ref{eq:loBound44}), we have\small
	\begin{equation}\label{eq:loBound5}
	\begin{split}
	&(M_f + 2c\sigma_{max}(\mathbf{W}))^2\|\mathbf{x}^{k+1} - \mathbf{x}^*\|_2^2 \\
	&\geq  2\langle c\mathbf{M}_+(\mathbf{z}^{k} - \mathbf{z}^{*}) - \mathbf{M}_-(\boldsymbol{\beta}^{k} - \boldsymbol{\beta}^*) , \frac{c}{2} (\mathbf{M}_+\mathbf{M}_+^T - \mathbf{M}_-\mathbf{M}_-^T)\mathbf{e}_x^k\rangle\\
	&+ \|c\mathbf{M}_+(\mathbf{z}^{k} - \mathbf{z}^{*}) - \mathbf{M}_-(\boldsymbol{\beta}^{k} - \boldsymbol{\beta}^*)\|_2^2\\
	& + \|\frac{c}{2} (\mathbf{M}_+\mathbf{M}_+^T - \mathbf{M}_-\mathbf{M}_-^T)\mathbf{e}_x^k\|_2^2.
	\end{split}
	\end{equation}
	\normalsize
	Taking average from both sides of (\ref{eq:loBound5}), and considering that the elements of $\mathbf{e}_x^k$ are zero mean random variables independent of $\mathbf{z}^k$ and $\boldsymbol{\beta}^k$, we have
	\begin{equation}\label{eq:loBound52}
	\begin{split}
	\mathbb{E}_{\mathcal{F}^{k+1}}&\big[\|\mathbf{x}^{k+1} - \mathbf{x}^*\|_2^2\big]\\
	&\geq \frac{c^2}{\big(M_f + 2c\sigma_{max}(\mathbf{W})\big)^2}\mathbb{E}\big[\|(\mathbf{L}_+ - \mathbf{L}_-)\mathbf{e}_x^k\|_2^2\big],
	\end{split}
	\end{equation}
	where $\mathbf{L}_+ = \frac{1}{2}\mathbf{M}_+\mathbf{M}_+^T$ and $\mathbf{L}_- = \frac{1}{2}\mathbf{M}_-\mathbf{M}_-^T$ are, respectively, the extended signless and signed Laplacian matrices of the underlying network. According to the structure of these matrices, the $(i,j)$ block of $\mathbf{P} = \mathbf{L}_+ - \mathbf{L}_-$ is as follows
	\begin{equation}\label{eq:diffDefin}
	\mathbf{P}(i ,j) = \left\{ \begin{array}{cc}
	\mathbf{0}_n & \hspace{1cm}  i = j, \text{or}\hspace{1mm} (i,j)\notin\mathcal{A}  \\
	
	2\mathbf{I}_n  & \hspace{1cm}  (i,j) \in \mathcal{A}
	\end{array} \right ..	
	\end{equation}
	According to \eqref{eq:diffDefin}, following some algebraic manipulations, we have
	\begin{equation}\label{eq:loBound6}
	\mathbb{E}_{\mathcal{F}^{k+1}}\big[\|\mathbf{x}^{k+1} - \mathbf{x}^*\|_2^2\big] \geq \frac{8nEc^2\sigma_n^2}{\big(M_f + 2c\sigma_{max}(\mathbf{W})\big)^2},
	\end{equation}
	which completes the proof.
%%%%%%%%%%%%%%
\section{Proof of Proposition 2}\label{ap:prop2}
According to the definition of matrices $\mathbf{A}$ and $\mathbf{B}$, the feasible set of problem in \eqref{matrixForm} is nonempty. This fact and the convexity of local objective functions guarantee that the Lagrangian function corresponding to \eqref{matrixForm} has a saddle point. If local objective functions are also proper and closed, inequality \eqref{eq:admmConvLy} in Proposition \ref{prep:boydConv} is valid for this problem. From \eqref{admmC}, we have
			\begin{equation}\label{eq:dualEqv}
			\mathbf{A}\mathbf{x}^{k+1} + \mathbf{B}\mathbf{z}^{k+1} = \frac{1}{c}\big(\boldsymbol{\lambda}^{k+1} - \boldsymbol{\lambda}^k\big).
			\end{equation}
			 Since $\boldsymbol{\lambda}^k = [\boldsymbol{\beta}^k;-\boldsymbol{\beta}^k]$, and $\mathbf{B} = [-\boldsymbol{I}_{2nE};-\boldsymbol{I}_{2nE}]$, and based on \eqref{eq:dualEqv}, \eqref{eq:admmConvLy} can be rewritten as follows
			\begin{equation}\label{eq:ourLyap}
			\begin{split}
			&\frac{1}{c}\|\boldsymbol{\beta}^{k+1} - \boldsymbol{\beta}^*\|_2^2 + c\|\mathbf{z}^{k+1} - \mathbf{z}^*\|_2^2\\
			&\leq \frac{1}{c}\|\boldsymbol{\beta}^{k} - \boldsymbol{\beta}^*\|_2^2 + c\|\mathbf{z}^{k} - \mathbf{z}^*\|_2^2\\
			&\hspace{5mm}  - \frac{1}{c}\|\boldsymbol{\beta}^{k+1} - \boldsymbol{\beta}^k\|_2^2 - c\|\mathbf{z}^{k+1} - \mathbf{z}^k\|_2^2,
			\end{split}
			\end{equation}
			and \eqref{eq:rZLimit} is equivalent to 	
			\begin{equation}\label{eq:rZLimit2}
			\underset{k \rightarrow \infty}{\lim}\hspace{1mm}\boldsymbol{\beta}^{k+1} - \boldsymbol{\beta}^k  = \mathbf{0}, \hspace{3mm} \underset{k \rightarrow \infty}{\lim}\hspace{1mm}\mathbf{z}^{k+1} - \mathbf{z}^k = \mathbf{0}.
			\end{equation}
			Replacing \eqref{eq:rZLimit2} in \eqref{midForm}, we have
			\begin{subequations}\label{eq:midForm2}
				\begin{equation}\label{eq:midFormA2}
				\underset{k \rightarrow \infty}{\lim}\hspace{1mm}\nabla f(\mathbf{x}^{k}) + \mathbf{M}_-\boldsymbol{\beta}^{k}  = \mathbf{0},
				\end{equation}
				\begin{equation}\label{eq:midFormB2}
					\underset{k \rightarrow \infty}{\lim}\hspace{1mm}\mathbf{M}_-^T\mathbf{x}^{k} = \mathbf{0},
				\end{equation}
				\begin{equation}\label{eq:midFormC2}
					\underset{k \rightarrow \infty}{\lim}\hspace{1mm}\frac{1}{2}\mathbf{M}_+^T\mathbf{x}^{k} - \mathbf{z}^k = \mathbf{0}.
				\end{equation}
			\end{subequations}
			Comparing \eqref{eq:midForm2} and \eqref{eq:KKTCond}, it can be seen that as $k \rightarrow \infty$, primal and dual variables satisfy KKT conditions, and hence the iteration converges to  an optimal point. In the other words,  \eqref{eq:rZLimit2} is a sufficient condition for  attaining an optimal point, which completes the proof.
%%%%%%%%%
\section{Proof of Corollary 1}\label{ap:Cor1}
From \eqref{eq:ourLyap} in Proposition \ref{prep:ramark3}, we have
		\begin{equation}\label{eq:boundTerm}
		\begin{split}
		& 0 \leq \frac{1}{c}\|\boldsymbol{\beta}^{k+1} - \boldsymbol{\beta}^k\|_2^2 + c\|\mathbf{z}^{k+1} - \mathbf{z}^k\|_2^2\\
		 & \hspace{5mm} \leq \frac{1}{c}\|\boldsymbol{\beta}^{k} - \boldsymbol{\beta}^*\|_2^2 + c\|\mathbf{z}^{k} - \mathbf{z}^*\|_2^2,
		 \end{split}
		\end{equation}	
		which is equivalent to
		\begin{equation}\label{eq:linearRelTerm}
		\begin{split}
		\frac{1}{c}&\|\boldsymbol{\beta}^{k+1} - \boldsymbol{\beta}^k\|_2^2 + c\|\mathbf{z}^{k+1} - \mathbf{z}^k\|_2^2\\
		 &= \eta^{k+1} \big(\frac{1}{c}\|\boldsymbol{\beta}^{k} - \boldsymbol{\beta}^*\|_2^2 + c\|\mathbf{z}^{k} - \mathbf{z}^*\|_2^2\big),
		 \end{split}
		\end{equation}
		where $0 < \eta^{k+1}\leq 1, \forall k \geq 0$. Note that, based on Preposition \ref{prep:ramark3}, the LHS of \eqref{eq:linearRelTerm} being zero,  guarantees that its RHS is also zero, and hence we can consider $0 < \eta^{k+1}$ in all cases. From \eqref{eq:ourLyap} and \eqref{eq:linearRelTerm}, we have
		\begin{equation}\label{eq:linearConvNonStr2}
		\begin{split}
		&\frac{1}{c}\|\boldsymbol{\beta}^{k+1} - \boldsymbol{\beta}^*\|_2^2 + c\|\mathbf{z}^{k+1} - \mathbf{z}^*\|_2^2\\
		&\leq \frac{1}{1 + \delta^{k+1}}\big(\frac{1}{c}\|\boldsymbol{\beta}^{k} - \boldsymbol{\beta}^*\|_2^2 + c\|\mathbf{z}^{k} - \mathbf{z}^*\|_2^2\big),
		\end{split}
		\end{equation}
		 where $1/(1+\delta^{k+1}) = 1 - \eta^{k+1}$, $\delta^0 = 0$ and $\delta^{k+1} > 0,\forall k\geq0$, which completes the proof.
%%%%%%%%%%%%%
\section{Proof of Theorem 4}\label{ap:Theorem4}
Similar to proof of Theorem \ref{linearConvTheo}, we have
{\color{black}
\begin{equation}\label{eq:nonStrProof}
\begin{split}
 \mathbb{E}_{\mathcal{F}^{k+1}}[d^{k+1}] &= \mathbb{E}_{\mathcal{F}^{k}}\big\{\mathbb{E}_{e^{k+1}}[d^{k+1}|\mathcal{F}^{k}]\big\} \\
 & \stackrel{\text{(a)}}{\leq} \frac{1}{1 + \delta^{k+1}}\mathbb{E}_{\mathcal{F}^k}[d^k] + 2cnE\sigma_n^2,
 \end{split}
\end{equation}}
where $(a)$ follows from \eqref{eq:linearConvNonStr} in Corollary \ref{cor:boundTermDec}. By recursive use of \eqref{eq:nonStrProof}, we have
{\color{black}
\begin{equation}\label{eq:nonStrP2}
  \mathbb{E}_{\mathcal{F}^{k+1}}[d^{k+1}] \leq \frac{d^0}{\prod_{\ell = 1}^{k+1}(1 + \delta^\ell)} + \sum_{ \ell = 0}^{k}\frac{1}{\prod_{i = 0}^{\ell}(1+\delta^i)}2cnE\sigma_n^2.
\end{equation}}
Since $\delta^k > 0, \forall k>0$, we have	
		\begin{equation}\label{eq:boundDkNonStr}
		\underset{k \rightarrow \infty}{\lim}\hspace{1mm} \mathbb{E}_{\mathcal{F}^{k}}[d^k] <\infty,
		\end{equation}
	and hence we almost surely have,
	\begin{equation}\label{eq:boundDkNonStr2}
	\underset{k \rightarrow \infty}{\lim}\hspace{1mm} d^k <\infty,
	\end{equation}
		which implies
		\begin{equation}\label{eq:betaZBound}
		\begin{split}
		&\underset{k \rightarrow \infty}{\lim}\hspace{1mm}\|\hat{\boldsymbol{\beta}}^{k+1} - \boldsymbol{\beta}^{*}\|_2 < \infty,\\
		&\underset{k \rightarrow \infty}{\lim}\hspace{1mm}\|\hat{\mathbf{z}}^{k+1} - \mathbf{z}^*\|_2 < \infty.
		\end{split}		
		\end{equation}
		From \eqref{noisyIterB2}, \eqref{noisyIterC2}, \eqref{eq:KKT2} and \eqref{eq:KKT3}, we have
		\begin{equation}\label{eq:xBoundNnStr}
			\begin{split}
			&\hat{\boldsymbol{\beta}}^{k+1} - \boldsymbol{\beta}^{*} = \hat{\boldsymbol{\beta}}^{k} - \boldsymbol{\beta}^{*} + \frac{c}{2}\mathbf{M}_-^T(\mathbf{x}^{k+1} - \mathbf{x}^*) + \frac{c}{2}\mathbf{M}_-^T\mathbf{e}_x^{k+1},\\			
			&\hat{\mathbf{z}}^{k+1} - \mathbf{z}^* =  \frac{1}{2}\mathbf{M}_+^T(\mathbf{x}^{k+1} - \mathbf{x}^*) + \frac{1}{2}\mathbf{M}_+^T\mathbf{e}_x^{k+1}.
			\end{split}		
		\end{equation}
		Since $\|\mathbf{e}_x^{k}\|_2 \leq e_{max} < \infty$, and based on the definition of $\mathbf{M}_-$ and $\mathbf{M}_+$, from \eqref{eq:betaZBound} and \eqref{eq:xBoundNnStr} we have
		\begin{equation*}
		\underset{k \rightarrow \infty}{\lim} \hspace{1mm} \|\mathbf{x}^k - \mathbf{x}^*\|_2^2 < \infty,
		\end{equation*}
		which completes the proof.	
%%%%%%%%%%%%
\bibliographystyle{IEEEtran}
\bibliography{refs}

\end{document}